\documentclass{article}

\usepackage{arxiv}

\usepackage[utf8]{inputenc} 
\usepackage[T1]{fontenc}    
\usepackage{hyperref}       
\usepackage{url}            
\usepackage{booktabs}       
\usepackage{amsfonts}       
\usepackage{nicefrac}       
\usepackage{microtype}      
\usepackage{lipsum}
\usepackage{graphicx}
\usepackage{subfigure}
\usepackage{amsmath}
\usepackage{caption}

\title{Bifrequency 3D Ghost Imaging with Haar Wavelet Transform}

\author{
  Mengjia Xi, Hui Chen\thanks{Corresponding author: chenhui@xjtu.edu.cn.}, Yuan Yuan, Gao Wang, Yuchen He, Jianbin Liu, Huaibin Zheng, Zhuo Xu\\
  Electronic Materials Research Laboratory,\\
  Key Laboratory of the Ministry of Education and International Center for Dielectric Research,\\
  School of Electronic and Information Engineering,\\
  School of Electronic and Information Engineering,\\
  Xi'an Jiaotong University, Xi'an 710049, China\\
   \And
 Yan Liang\\
  Institute of Wide Band Gap Semiconductors,\\
  Shaanxi Key Lab of Information Photonic Technique,\\
  School of Electronic and Information Engineering,\\
  Xi'an Jiaotong University, Xi'an 710049, China\\
}

\begin{document}
\maketitle

\begin{abstract}

Recently, ghost imaging has been attracting attentions because its mechanism would lead to many applications inaccessible to conventional imaging methods. However, it is challenging for high contrast and high resolution imaging, due to its low signal-to-noise ratio (SNR) and the demand of high sampling rate in detection.
To circumvent these challenges, we here propose a ghost imaging scheme that exploits Haar wavelets as illuminating patterns with a bi-frequency light projecting system and frequency-selecting single-pixel detectors.
This method provides a theoretically 100\% image contrast and high detection SNR, which reduces the requirement of high dynamic range of detectors, enabling high resolution ghost imaging. Moreover, it can highly reduce the sampling rate (far below Nyquist limit) for a sparse object by adaptively abandoning unnecessary patterns during the measurement. These characteristics are experimentally verified with a resolution of $512\times512$ and a sampling rate lower than 5\%.
A high-resolution ($1000\times1000\times1000$) 3D reconstruction of an object is also achieved from multi-angle images.

\end{abstract}

\section{Introduction}
\label{sec:Introduction}

Ghost imaging (GI) has been attracted more and more attentions owing to its unique imaging technique\cite{pittman1995optical,valencia2005two-photon,zhang2005correlated}. Compared with traditional optical imaging techniques, GI illuminates a scene with a sequence of optical patterns and uses a bucket detector to collect the total reflected or transmitted light from the object. The correlation between the bucket signal and the patterns reveals the image of the object.
Computational ghost imaging (CGI) was later proposed\cite{shapiro2009computational} and experimentally demonstrated soon after\cite{bromberg2009ghost}, providing a practical scheme for the applications of GI.

Without recourse to imaging lenses or spatial resolving detectors, GI possesses many advantages, such as lensless imaging capability\cite{scarcelli2006can}, high detection sensitivity\cite{morris2015imaging}, and being applicable to scenarios lacking array detectors\cite{Peng2015The,Liu2017Computational,Aspden2015Photon}. GI therefore has a lot potential applications in different fields from optical imaging\cite{gong2016three}, X-ray imaging\cite{pelliccia2016experimental,yu2016fourier,zhang2018tabletop}, biological diagnostics\cite{ota2018ghost} to atomic sensing\cite{khakimov2016ghost,baldwin2017ghost}.

However, GI is still challenging for high-resolution imaging and high SNR detection. The imaging resolution is subject to the size of speckles projected onto a scene. A smaller size of the speckles yields a higher resolution, causing more speckles emerging within the scene. Consequently, the mean value of the bucket signal will increase and its variation becomes relatively small. As the resolution increases, the variance of the bucket signal is harder and harder to be detected, demanding a very high dynamic range of single-pixel detectors. Moreover, for an $N$-pixel image (the average number of speckles on the scene is $N$), GI requires at least $N$ samples to recover the image, so called the Nyquist limit of the measurement. Especially, GI with the random speckles usually needs much more samples than $N$. Thus, for high-resolution imaging, a large samples will cause low imaging speed, standing in the way of the application of ghost imaging.

To overcome these challenges, we introduce the wavelet transform (WT)\cite{mallat1989a,daubechies1992ten} into the light source modulation of the CGI system. The speckle patterns are designed based on a modified 2D Haar wavelets\cite{haar1911zur}, which are projected onto an object with a bi-frequency illuminating source. The reflected light from the object is measured with two frequency-selecting single-pixel detectors. The image is recovered based on the theory of inverse wavelet transform (IWT). We term this method "Bifrequency Wavelet Ghost Imaging (BiWave-GI)", which has the following three advantages. (1) It provides a principally 100\% image contrast. Experimentally, we can easily acquired a contrast more than 90\% under a resolution of $512\times512$. (2) BiWave-GI detects an object from a low resolution to a high resolution. On each resolution level, it performs a block (the size is determined by the resolution) scanning on the object. This results in a very high detection SNR, which dramatically reduces the requirement of high dynamic range of detectors. Even with 1-bit detectors (only differentiate on or off), BiWave-GI can still provide high-contract and high-resolution results. (3) On each resolution level, we adaptively abandon unnecessary samples based on the detection of the previous lower resolution level. Thus, BiWave-GI can measure a spare object far below the Nyquist limit without any prior knowledge of the object or any iterative reconstruction algorithm, which is quite different than the methods using compressive sensing\cite{gong_sparse_2015}. We experimentally demonstrate a high-resolution imaging with a sampling rate lower than 5\%. These advantages
enables us easily and efficiently obtain high-contrast and high-resolution images with CGI, which also paves a way for 3D ghost imaging with high-resolution. Experimentally, we combine BiWave-GI and space carving algorithm\cite{szeliski1993rapid,laurentini1995far,kutulakos2000theory} together and reconstruct a 3D imaging with $1000\times1000\times1000$ resolution.

Recently, wavelet analysis has been considered as an important tool for ghost imaging and single-pixel imaging\cite{amann_compressive_2013,yu2014adaptive,rousset2017adaptive,alemohammad2017high-speed}. Some researches focused on using wavelet analysis to post-process images rather than using wavelet bases as 2D illuminating patterns\cite{amann_compressive_2013,yu2014adaptive,rousset2017adaptive}, which did not take the advantage of high detection SNR as we acquired in our experiments. There is a work that used 1D Haar wavelet patterns to illuminate a moving object and obtained fast imaging\cite{alemohammad2017high-speed}. This previous work\cite{alemohammad2017high-speed} could not implement low sampling rate detection, because it lacks 2D wavelet analysis during the measurement. Our method provides a comprehensive solution to the challenges including high-contrast, high-resolution and adaptive low sampling rate, which would make a big step to the application of ghost imaging.

\section{Principle}
\label{sec:Principle}

Wavelets basis functions are mathematical functions that maps data onto different frequency components, and then study each component with a resolution matched to its scale.
They have advantages over traditional Fourier methods in analyzing physical situations where the signal contains discontinuities and sharp spikes\cite{strang1993wavelet}. We here select Haar wavelets\cite{strang1996wavelets} (one of the simplest wavelets), whose mother wavelet function is a binary function:
\begin{eqnarray}
\varphi (t)=\begin{cases} \ \ \ 1,&t\in [0,\frac{1}{2}] \cr {-1},&t\in [\frac{1}{2},1] \cr \ \ \ 0,&otherwise\end{cases}.
\label{eq:ref1}
\end{eqnarray}
The Haar wavelet basis functions (daughter wavelets) is constructed from the mother wavelet by scaling and shifting operation. For a $N$-pixel Haar transform, the $j$-th daughter wavelet is written as:
\begin{eqnarray}
H_j=\sqrt{2^{s-q}}\varphi(2^{s-q}\cdot t-k),
\label{eq:ref2}
\end{eqnarray}
where $q=log_2N$. $s=0,1,2,...,q-1$, denoting the scaling level. $k=0,1,...,2^j-1$, representing the shifting factor. $j=2^{s+1}-1+k$, indicating the $j$-th basis. We then reshape each daughter wavelet to a $n\times n$ ($n=\sqrt{N}$) 2D matrix. There are many different type of 2D transformations. Here, we choose the simplest one that is directly dividing a basis into $n$ segments, each of which becomes a row of the 2D matrix. It can be formulated as below:
\begin{eqnarray}
	M_j(x,y)=\sqrt{2^{s-q}}\varphi\left(2^{s-q}\cdot ((y-1)n+x)-k\right).
	\label{eq:SimpleH2D}
\end{eqnarray}
In the subsection of ``Low Sampling Rate'', we will show another type of 2D matrices, whose transformation is a little complicated, but which has a lower sampling rate and is more suitable for 2D images .

\begin{figure}[htbp]
	\centering
	\includegraphics[width=15cm]{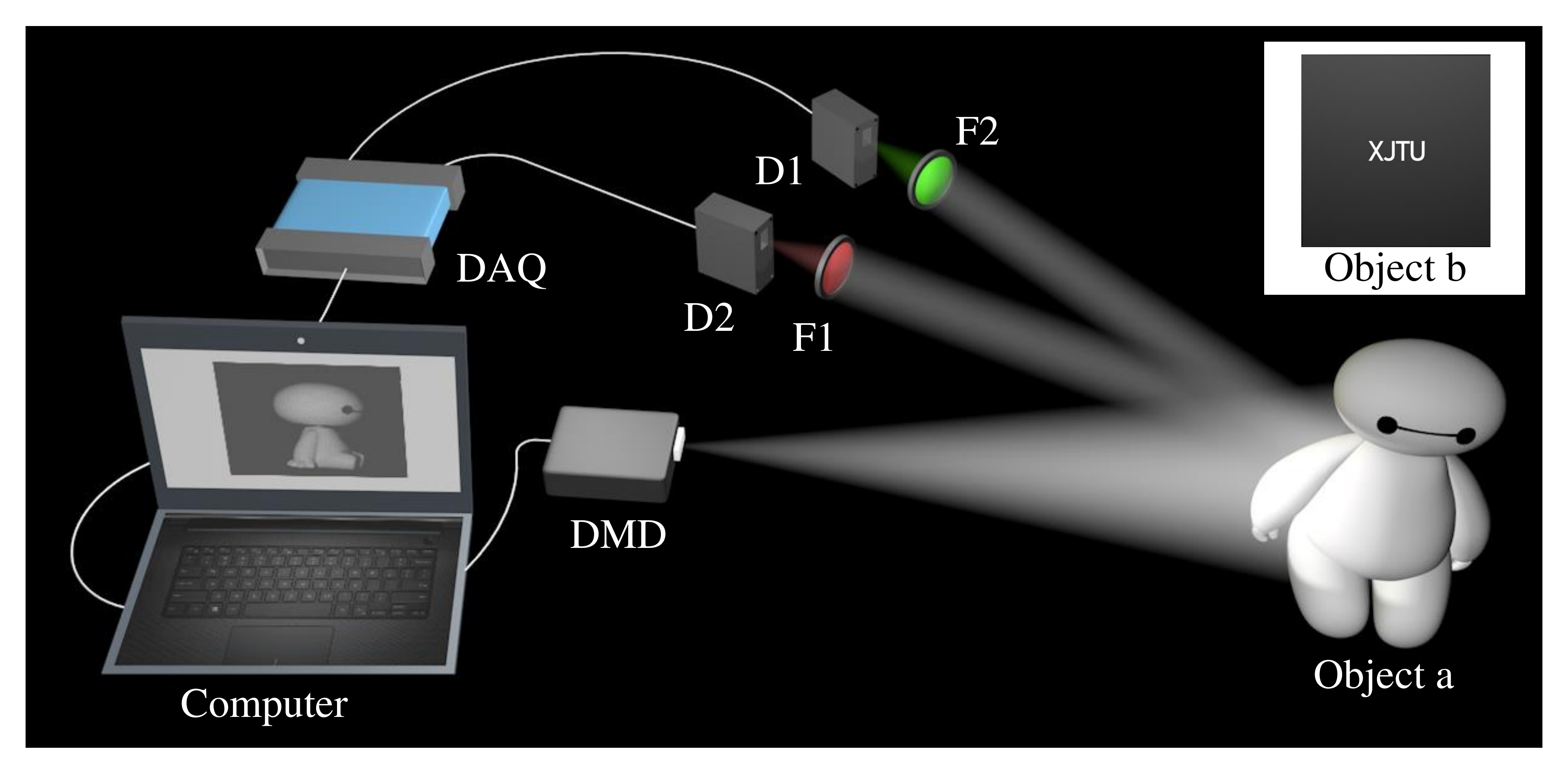}
	\caption{The schematic of the experimental setup. The light sources is a DMD. F1 and F2 are red and green filters; D1 and D2 are bucket detectors. We tested the imager with a 3D object (a Baymax model) and a sparse object (letters of "XJTU"). }
	\label{fig:1}
\end{figure}
A schematic of the experimental setup is shown in Fig. \ref{fig:1}. A sequence of the 2D matrices are projected onto an object with a DMD (digital micro-mirror device). The resolution is $512\times512$ pixels. Since the Haar wavelets consist of $+1$, $0$ and $-1$ in mathematics. We therefore use two frequencies of light to represent $+1$ and $-1$, respectively. In the experiments, we use DMD's embedded red and green light sources to implement the bi-frequency configuration. Correspondingly, the reflected light from the objects are measured with two single-pixel detectors. Red and green band-pass filters are placed in front of the two detectors respectively, making frequency-selecting detectors. When the object is illuminated with $M_j$, the two detectors respectively measure the intensities from two different-frequencies light, and gives two bucket signals denoting as $I^{1}_j$ and $I^{2}_j$. The total bucket signal is constructed as $B_j=I^{1}_j-I^{2}_j$. After projecting all set of Modified Haar 2D matrices onto the object, we obtain a set of bucket signals:
\begin{eqnarray}
&&\{M_1,M_2,...,M_N\}\cdot {\bf O}bj=\{B_1,B_2,...,B_N\}\cr
&&\equiv {\bf M} \cdot Obj={\bf B}.
\end{eqnarray}
The image is then reconstructed by inverse wavelet transform:
\begin{equation}
	{\bf O}bj = {\bf M}^{-1} \cdot {\bf B}.
\end{equation}
Theoretically, the reconstruction has zero background, which is different than the ghost imaging with random speckle patterns.

Multi-resolution (or multi-scale) analysis is an important characteristics of Haar wavelets. As described by Eq. (\ref{eq:SimpleH2D}), in each scaling level, 2D Haar wavelets acts as using two light strips (1 and -1) to scan (by the shifting operation) an object, and reconstruct an image at this resolution level. This scanning operation causes a high detection SNR for each scaling level. Thus, it dramatically reduces the requirement of the high dynamic range of detectors for ghost imaging.

In the following, we will experimentally show that this imager provides an almost background-free image with a high resolution and very low dynamic range of detectors, as well as a low sampling rate for a sparse object.

\section{Experiment Results and Discussions}
\label{sec:Experiment Results and Discussions}

\begin{figure}[htbp]
	\centering
	\includegraphics[width=10cm]{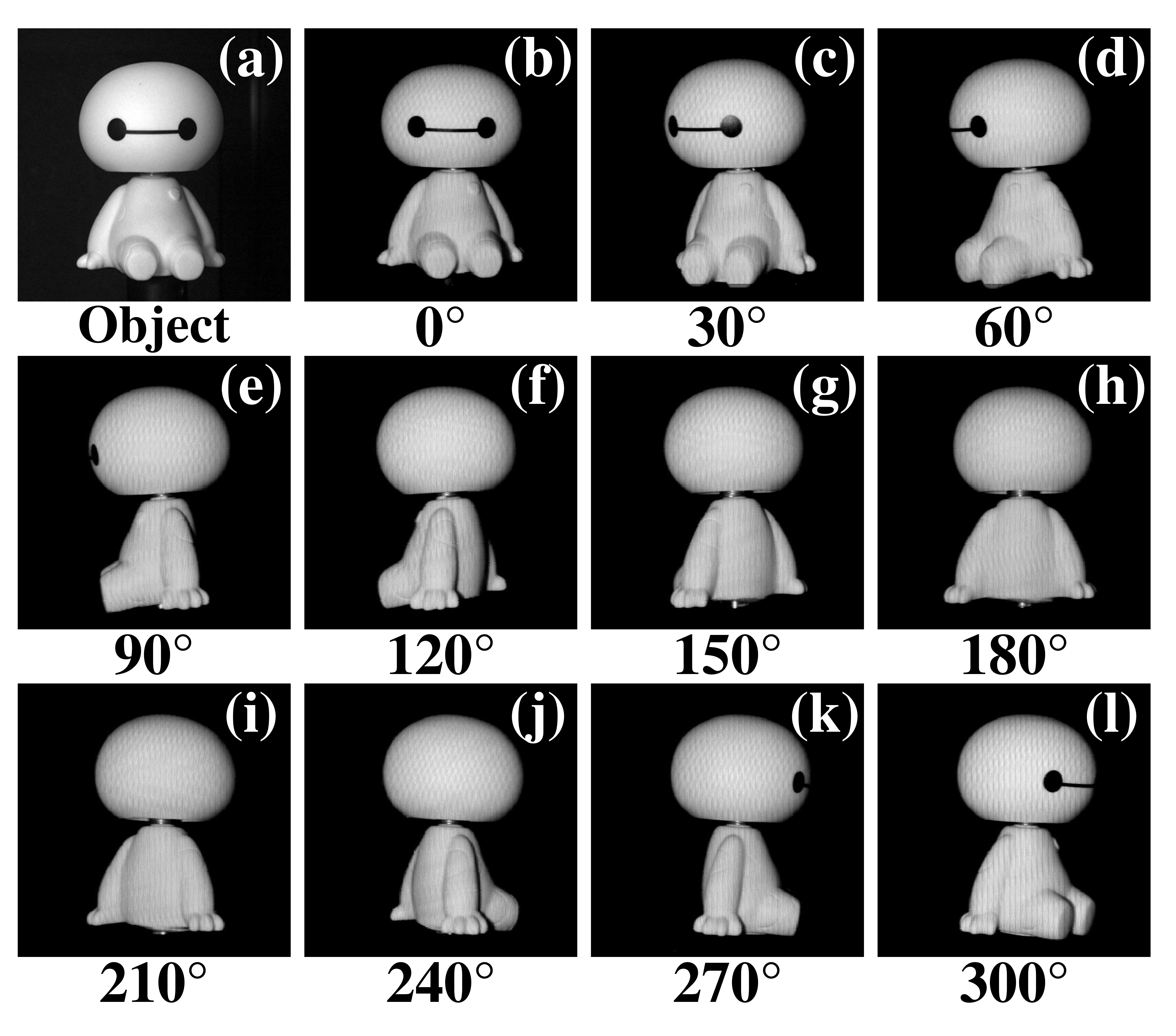}
	\caption{ The images of the Baymax at multiple angles using Biwave-GI with a resolution of $512\times512$.}
	\label{fig:3}
\end{figure}
In the experiment, we image the Baymax model shown in Fig. \ref{fig:3}(a) using the  modified Haar wavelet patterns \{$M_j$\} with a $512\times512$ resolution. The experimental results are shown in Fig. \ref{fig:3}. As predicted by the above theory, the background of the images are almost zeros. Note that, we did not use any imaging enhancement process to remove the background. Since we are able to capture high-contrast and high-resolution images with different angles, a 3D image can be acquired with active passive 3D methods such as space carving algorithm\cite{szeliski1993rapid,laurentini1995far,kutulakos2000theory}.

\begin{figure}[htbp]
	\centering
	\includegraphics[width=7cm]{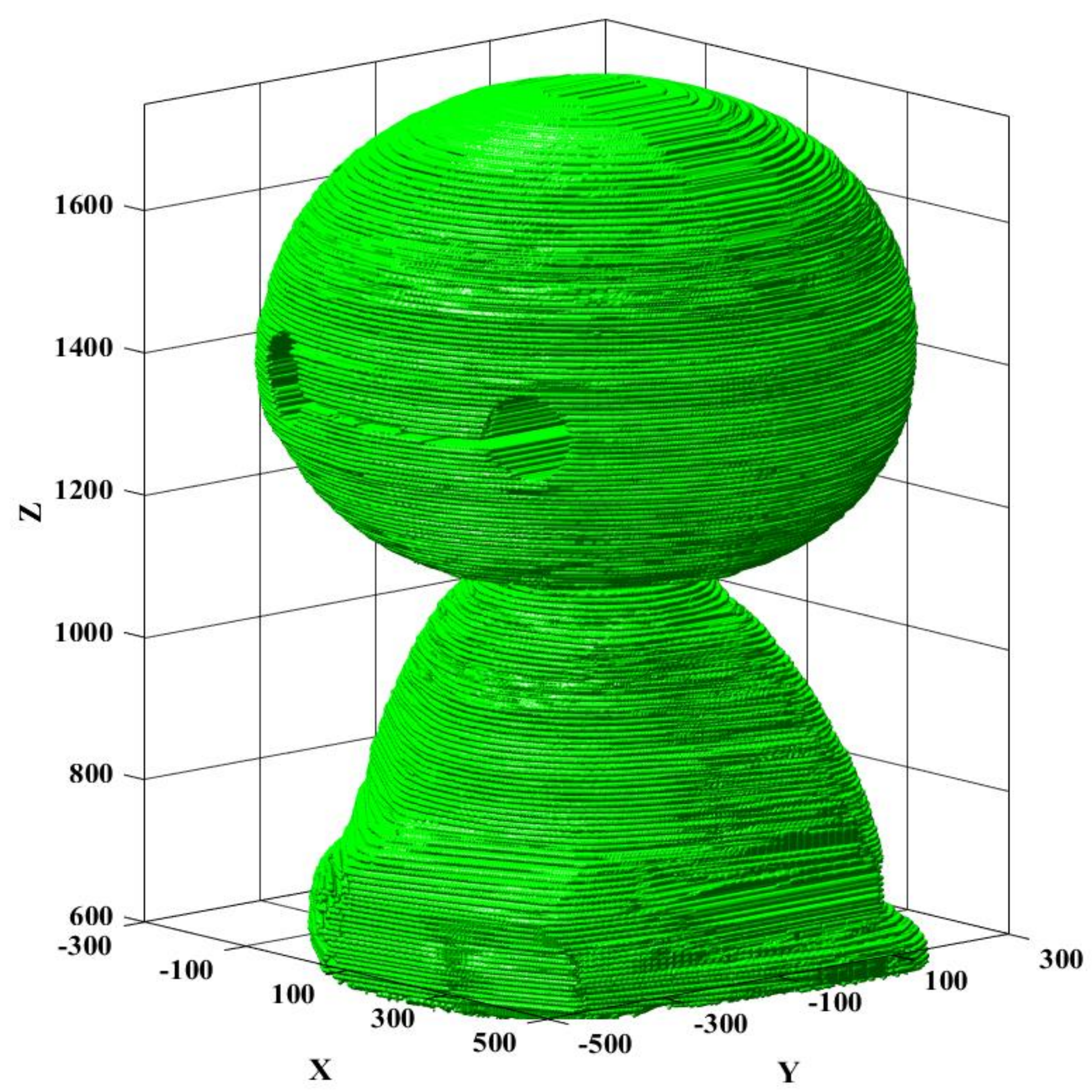}
	\caption{The 3D imaging result of the Baymax with space carving algorithm. The resolution is $1000\times 1000\times 1000$.}
	\label{fig:4}
\end{figure}

We first generated a big cube with a scale of $2048\times 2048\times 2048$ and divided it into $1000\times 1000\times 1000$ small cubes, each small cube with a resolution of 2.048.
A total of 72 images with different angles (5 degrees apart) were obtained experimentally. Then, the image result in a certain angle is binarized, and the silhouette of the object is extracted. Based on the silhouette result, the big cube is engraved along the direction, the small cubes without object are removed.
The images of each angle are processed according to this method, and the results of the big cube engraving at each angle are intersected to obtain the 3D reconstruction result with $1000\times 1000\times 1000$ resolution shown in Fig. \ref{fig:4}

\subsection{Low Dynamic Range Requirements for Detectors}

In a traditional ghost imaging using random patterns, if the detectors has an infinity high dynamic range, background can be removed by post processing.
However, every detector has a finite dynamic range. When the required resolution (number of pixels) gets higher, the background of the bucket signal increases. Then, the detector's dynamic range rapidly becomes insufficient to distinguish the variance of the signal. For example, if the signal of the bucket detector has a mean value of 100 (i.e. $\bar{B}=100$) with a standard deviation of $\Delta B=10$, a detector with a dynamic range worse than 10 is unable to distinguish the variance of the signal, which will result in the failure of ghost imaging. The more accuracy is required to detect the variance of the bucket signal, the better reconstruction of the image, and the higher dynamic range is demanded. Moreover, the required dynamic range increases along with the growth of $\bar{B}$ that is proportional to the number of the speckles transmitted through or reflected from the object. Therefore, this instinctive scheme prevents ghost imaging from high-resolution.

In contrast, as analyzed in last section, Biwave-GI can dramatically reduce the requirement of the dynamic range. Even when the dynamic range is only 2, Biwave-GI can still recover a clear image. For a comparison purpose, we carry out computational ghost imaging experiments with random speckles (RCGI) and Hardamard Speckle (HCGI) respectively, and process the data at different dynamic ranges. The qualities of the recovered images are evaluated by the Structural Similarity Index (SSIM),defined as:
\begin{eqnarray}
SSIM(x,y)&\equiv& \frac{(2\mu_x\mu_y+C_1)(2\sigma_{xy}+C_2)}{(\mu_x^2+\mu_y^2+C_1)(\sigma_x^2+\sigma_y^2+C_2)},
\label{eq:ref5}
\end{eqnarray}

Where $\mu_x$, $\mu_y$, $\sigma_x$,$\sigma_y$, and $\sigma_{xy}$ are the local means, standard deviations, and cross-covariance for images x, y.
And $C1 = (0.01*L)^2$, $C2 = (0.03*L)^2$, $C3=C2/2$; where L is the specified dynamic range value.
To ensure a fair comparison, we reconstructed the image with the same resolution and the same number of speckles for the three methods: for a resolution of $512\times 512$, the maximum number of speckles illuminating the scene is $512\times 512=262144$.

\begin{figure}[htbp]
	\centering
	\includegraphics[width=15cm]{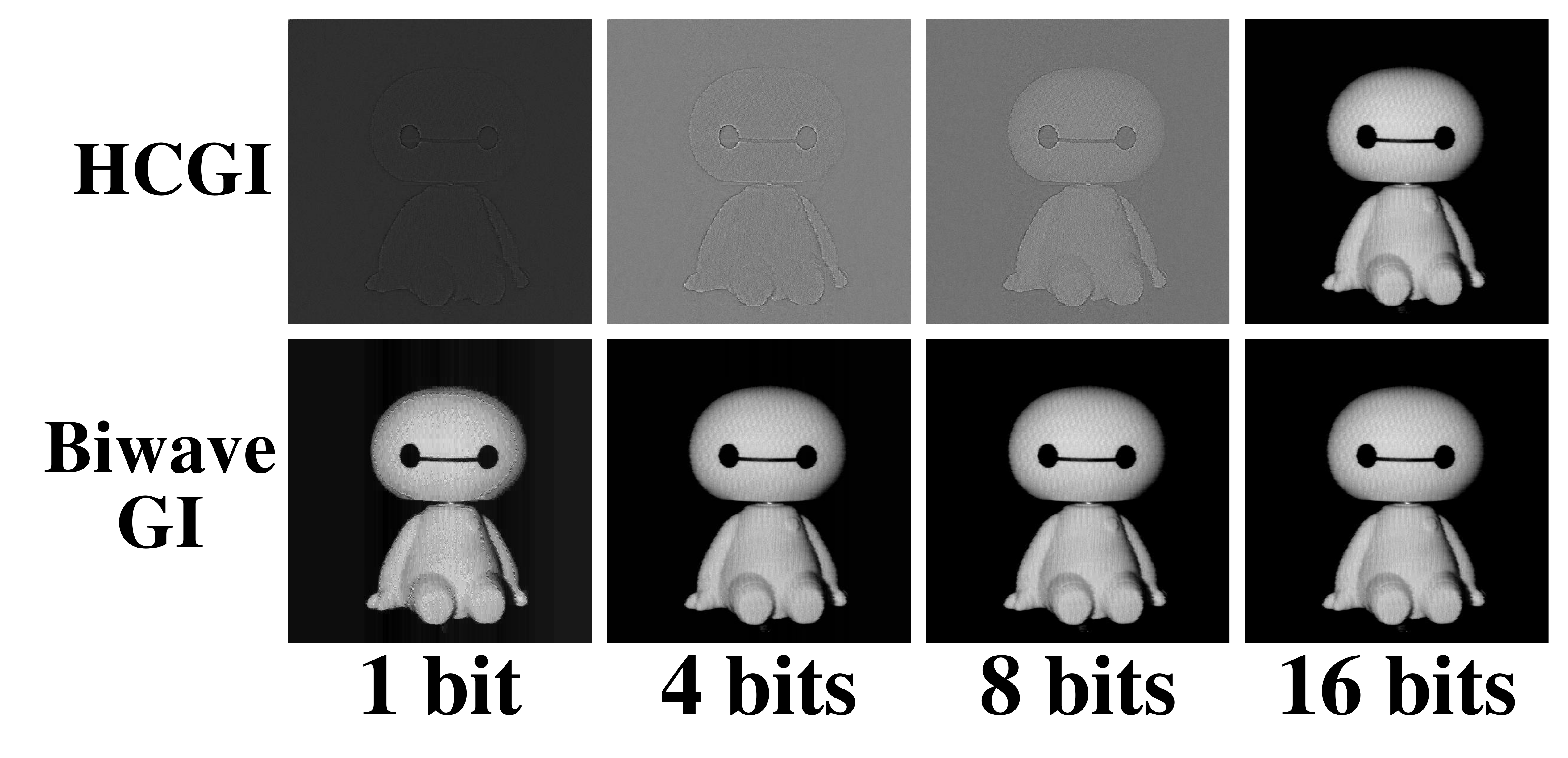}
	\caption{Experiment results of Biwave-GI, HCGI under different detector dynamic ranges (1,\ 4,\ 8 and 16 bits).}
	\label{fig:6}
\end{figure}

Figure \ref{fig:6} shows images reconstructed with BiWave-GI and HCGI, where the detector dynamic ranges were changed from $2$(1 bit) to $65536$ (16 bits).
Since RCGI failed to recover an image of a $64\times 64$ resolution under a dynamic range of 16 bits, the results of RCGI is not shown on the figure.

\begin{figure}[htbp]
	\centering
	\includegraphics[width=10cm]{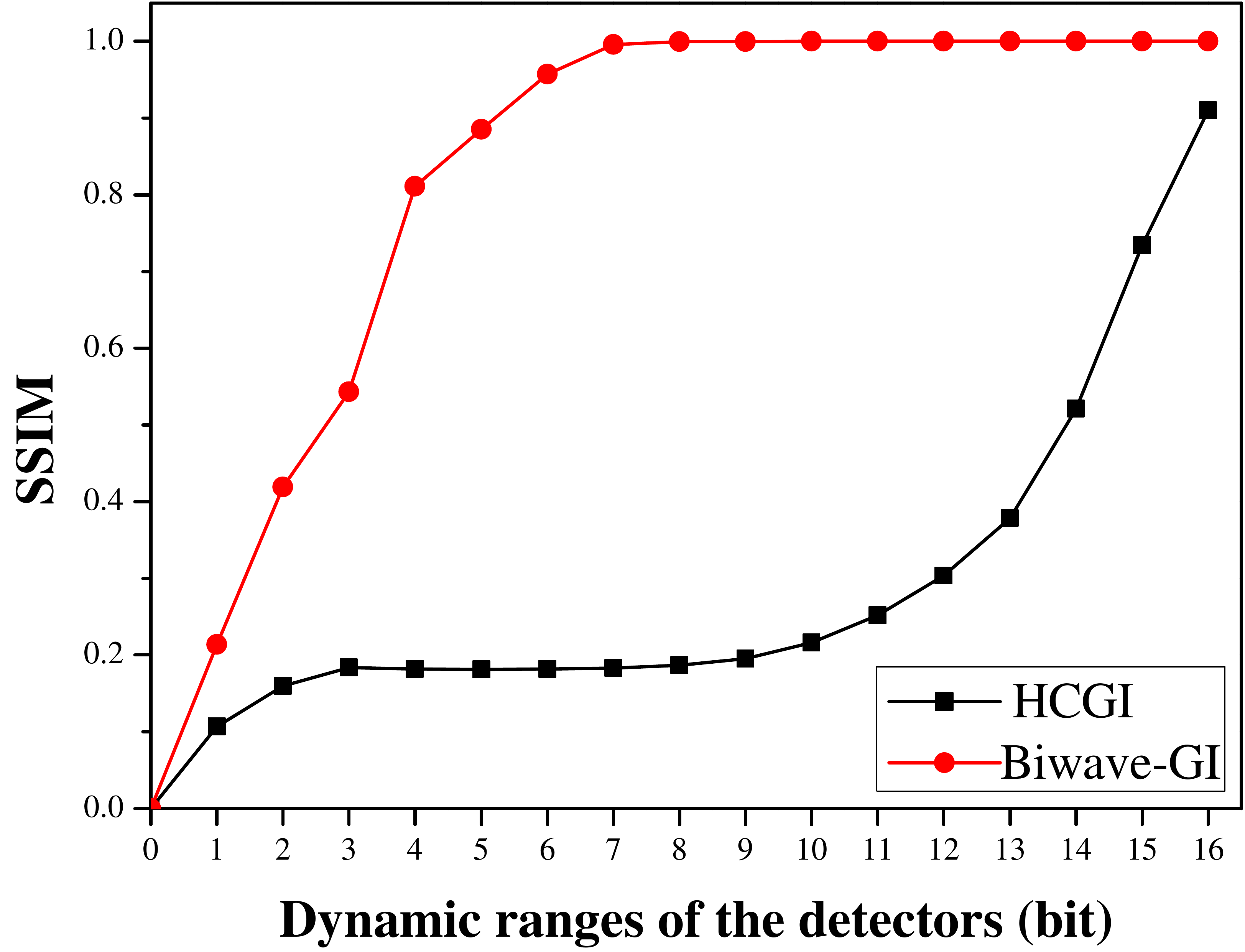}
	\caption{SSIM value curves of Biwave-GI and HCGI under different detector dynamic ranges (1 to 16 bits), respectively.}
	\label{fig:7}
\end{figure}

For a more quantitative comparison, a plot of dynamic ranges versus SSIM for each method was shown in Fig. \ref{fig:7}, where each curve corresponds to one result shown in Fig. \ref{fig:6}. One can see that Biwave-GI has high SSIM with relatively low dynamic ranges.

Biwave-GI requires far less dynamic ranges than the other two imaging methods.

\subsection{Low Sampling Rate}
\label{sec:Low Sampling Rate}

Another challenge of ghost imaging is that the number of the required illuminating patterns increases along with the increasing of the required resolution. For orthogonal patterns such as Hadamard speckles, the number of the patterns is equal to the resolution. For example, the resolution of $512\times 512$ requires  $262144$ patterns (the sampling rate is 100\%). If random speckles are used, much more patterns are needed (the sampling rate $\gg 100\%$). This would result in very long time to project a large number of patterns, limiting the application of ghost imaging with high resolutions. Although the methods using compressed sensing can reduce the sampling rate, they require prior knowledge of spare objects and time-consuming calculations. We here demonstrate that Biwave-GI can highly reduce the sampling rate for a spare object without any prior knowledge or complicated calculations. Because Haar wavelet can analyze signals from low to high scaling level (resolution) gradually, we design a self-adaptive projecting scheme that reduces the number of patterns for the next higher resolution based on the last reconstruction of the lower resolution.

In the experiment, we group the wavelet bases at the same scaling level as a cluster.
For a $512\times 512$ resolution, there are $19$ clusters, and their corresponding resolutions are getting higher and higher. We project patterns on an object cluster by cluster. On each scaling level, we recover an image at this resolution level, from which we can tell which bases of the next scaling level would result in zero or very small bucket signal. We then remove those bases from the next cluster, reducing the number of the projected patterns.

\begin{figure}[htbp]
	\centering
	\includegraphics[width=15cm]{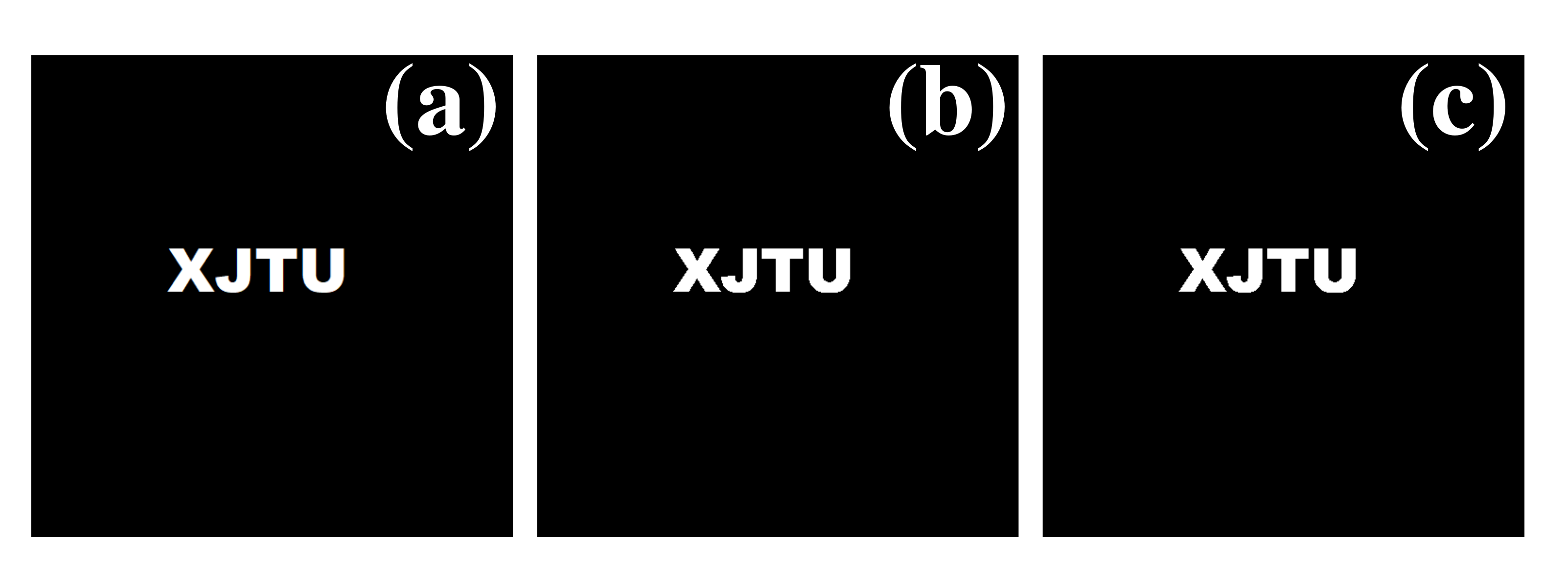}
	\caption{(a) The orginal object. (b) The recovered image of `XJTU' at a sampling rate of $4.8\%$ with {$M_j$}. (c) The recovered image of `XJTU' at a sampling rate of $2.4\%$ with {$Q_j$}.}
	\label{fig:8}
\end{figure}

Figure \ref{fig:8}(b) shows the experiment results of Biwave-GI that only uses 12614 patterns (the sampling rate is $4.8\%$) for
a spare object with letters of "XJTU". Note that, the duty ration (object's area/scene's area) is 1.5\%, which directly affects the sampling rate. The smaller is the duty ratio, the lower would be the sampling rate. In practical, a target in the sky has a very small duty ratio, which would be a feasible scenario for applying this imaging method.

Furthermore, different transformation from 1D Harr wavelets to 2D matrices would affect the sampling rate. We here test a quadratic blocking transformation: at lowest scaling level, the 2D scene is divided into four blocks (top-left,top-right,bottom-left,bottom-right); at the next level, each block is divided into four sub-blocks; this dividing is repeated until the highest scaling is reached. To represent this type of 2D wavelets, its daughter wavelet can be formulated as
\begin{eqnarray}
\varphi (x,y)=\begin{cases} \ \ \ 1,&x\in [0,\frac{1}{2}]\cap y\in [0,\frac{1}{2}]\cr
{-1},&x\in [\frac{1}{2},1] \cap y\in [0,\frac{1}{2}] \cr \ \ \ 0,&otherwise\end{cases}.
\label{eq:T2daughter}
\end{eqnarray}
A $n\times n$ 2D wavelet basis function can be constructed from the daughter wavelet:
\begin{eqnarray}
Q_j(x,y)=\sqrt{2^{s-L}}\varphi(2^{s-L}\cdot x-\alpha,2^{s-L}\cdot y-\beta/2),
\label{eq:QuadH2D}
\end{eqnarray}
where $L=log_2 n$. $s=0,1,2,...,L-1$, denoting the scaling level.
$\alpha=0,1,...,2^j-1$ and $\beta=0,1,...,2^{j+1}-1$, representing the shifting factors along x and y axes respectively. $j=(4^{s+1}-1)/3+2^\alpha+\beta$, indicating the $j$-th base.
After {$Q_j(x,y)$} are used as illuminating patterns, the sampling rate to recover the same object can be reduced to $2.4\%$, as shown in Fig.\ref{fig:8}(c).

\section{Conclusion}
\label{sec:Conclusion}

We have proposed and demonstrated a method exploiting a modified Haar waveles as illuminating patterns for ghost imaging, which enables background-free high resolution imaging. The method also greatly reduces the requirement of the dynamic range of bucket detectors.
For a sparse object, the imager can recover the image with a very low sampling rate, expediting the imaging speed. Furthermore, if the prior knowledge of a sparse object can be obtained, a special transformation $T$ could be designed to further reduce the sampling rate. These advantages pays a way to the application of ghost imaging.

\section*{Funding Information}

National Natural Science Foundation of China under Grant 11503020.

\bibliographystyle{unsrt}  

\bibliography{main}

\begin{thebibliography}{10}

\bibitem{pittman1995optical}
T~B Pittman, Yanhua Shih, Dmitry Strekalov, and Alexander~V Sergienko.
\newblock Optical imaging by means of two-photon quantum entanglement.
\newblock {\em Physical Review A}, 52(5), 1995.

\bibitem{valencia2005two-photon}
Alejandra Valencia, Giuliano Scarcelli, Milena Dangelo, and Yanhua Shih.
\newblock Two-photon imaging with thermal light.
\newblock {\em Physical Review Letters}, 94(6):063601, 2005.

\bibitem{zhang2005correlated}
Da~Zhang, Yanhua Zhai, Lingan Wu, and Xihao Chen.
\newblock Correlated two-photon imaging with true thermal light.
\newblock {\em Optics Letters}, 30(18):2354--2356, 2005.

\bibitem{shapiro2009computational}
Jeffrey~H Shapiro.
\newblock Computational ghost imaging.
\newblock {\em Physical Review A}, 78(6):061802, 2008.

\bibitem{bromberg2009ghost}
Yaron Bromberg, Ori Katz, and Yaron Silberberg.
\newblock Ghost imaging with a single detector.
\newblock {\em Physical Review A}, 79(5), 2009.

\bibitem{scarcelli2006can}
Giuliano Scarcelli, Vincenzo Berardi, and Yanhua Shih.
\newblock Can two-photon correlation of chaotic light be considered as
  correlation of intensity fluctuations?
\newblock {\em Phys. Rev. Lett.}, 96(6):063602, 2006.

\bibitem{morris2015imaging}
Peter~A Morris, Reuben~S Aspden, Jessica~EC Bell, Robert~W Boyd, and Miles~J
  Padgett.
\newblock Imaging with a small number of photons.
\newblock {\em Nat. Commun.}, 6:5913, 2015.

\bibitem{Peng2015The}
Hongtao Peng, Zhaohua Yang, Dapeng Li, and Ling~An Wu.
\newblock The application of ghost imaging in infrared imaging detection
  technology.
\newblock In {\em Selected Papers of the Photoelectronic Technology Committee
  Conferences Held June–july}, 2015.

\bibitem{Liu2017Computational}
Hong~Chao Liu and Shuang Zhang.
\newblock Computational ghost imaging of hot objects in long-wave infrared
  range.
\newblock {\em Applied Physics Letters}, 111(3):27--224, 2017.

\bibitem{Aspden2015Photon}
Reuben~S. Aspden, Nathan~R. Gemmell, Peter~A. Morris, Daniel~S. Tasca, Lena
  Mertens, Michael~G. Tanner, Robert~A. Kirkwood, Alessandro Ruggeri, Alberto
  Tosi, and Robert~W. Boyd.
\newblock Photon-sparse microscopy: visible light imaging using infrared
  illumination.
\newblock {\em Optica}, 2(12):1049, 2015.

\bibitem{gong2016three}
Wenlin Gong, Chengqiang Zhao, Hong Yu, Mingliang Chen, Wendong Xu, and
  Shensheng Han.
\newblock Three-dimensional ghost imaging lidar via sparsity constraint.
\newblock {\em Sci. Rep.}, 6:26133, 2016.

\bibitem{pelliccia2016experimental}
Daniele Pelliccia, Alexander Rack, Mario Scheel, Valentina Cantelli, and
  David~M Paganin.
\newblock Experimental x-ray ghost imaging.
\newblock {\em Phys. Rev. Lett.}, 117(11):113902, 2016.

\bibitem{yu2016fourier}
Hong Yu, Ronghua Lu, Shensheng Han, Honglan Xie, Guohao Du, Tiqiao Xiao, and
  Daming Zhu.
\newblock Fourier-transform ghost imaging with hard x rays.
\newblock {\em Phys. Rev. Lett.}, 117(11):113901, 2016.

\bibitem{zhang2018tabletop}
Ai-Xin Zhang, Yu-Hang He, Ling-An Wu, Li-Ming Chen, and Bing-Bing Wang.
\newblock Tabletop x-ray ghost imaging with ultra-low radiation.
\newblock {\em Optica}, 5(4):374--377, 2018.

\bibitem{ota2018ghost}
Sadao Ota, Ryoichi Horisaki, Yoko Kawamura, Masashi Ugawa, Issei Sato, Kazuki
  Hashimoto, Ryosuke Kamesawa, Kotaro Setoyama, Satoko Yamaguchi, Katsuhito
  Fujiu, et~al.
\newblock Ghost cytometry.
\newblock {\em Science}, 360(6394):1246--1251, 2018.

\bibitem{khakimov2016ghost}
Roman~I Khakimov, BM~Henson, DK~Shin, SS~Hodgman, RG~Dall, KGH Baldwin, and
  AG~Truscott.
\newblock Ghost imaging with atoms.
\newblock {\em Nature}, 540(7631):100, 2016.

\bibitem{baldwin2017ghost}
KGH Baldwin, RI~Khakimov, BM~Henson, DK~Shin, SS~Hodgman, RG~Dall, and
  AG~Truscott.
\newblock Ghost imaging with atoms and photons for remote sensing.
\newblock In {\em Light, Energy and the Environment}, page EM4B.1. Optical
  Society of America, 2017.

\bibitem{mallat1989a}
Stephane Mallat.
\newblock A theory for multiresolution signal decomposition: the wavelet
  representation.
\newblock {\em IEEE Transactions on Pattern Analysis and Machine Intelligence},
  11(7):674--693, 1989.

\bibitem{daubechies1992ten}
Ingrid Daubechies and Christopher Heil.
\newblock Ten lectures on wavelets.
\newblock {\em Computers in Physics}, 6(6):697--697, 1992.

\bibitem{haar1911zur}
Alfred Haar.
\newblock Zur theorie der orthogonalen funktionensysteme.
\newblock {\em Mathematische Annalen}, 71(1):38--53, 1911.

\bibitem{gong_sparse_2015}
Wenlin Gong and Shensheng Han.
\newblock High-resolution far-field ghost imaging via sparsity constraint.
\newblock {\em Sci. Rep.}, 5:9280, 2015.

\bibitem{szeliski1993rapid}
Richard Szeliski.
\newblock Rapid octree construction from image sequences.
\newblock {\em CVGIP: Image understanding}, 58(1):23--32, 1993.

\bibitem{laurentini1995far}
Aldo Laurentini.
\newblock How far 3d shapes can be understood from 2d silhouettes.
\newblock {\em IEEE Trans. Pattern Analysis and Machine Intelligence},
  17(2):188--195, 1995.

\bibitem{kutulakos2000theory}
Kiriakos~N Kutulakos and Steven~M Seitz.
\newblock A theory of shape by space carving.
\newblock {\em Int. J. Comput. Vis.}, 38(3):199--218, 2000.

\bibitem{amann_compressive_2013}
Marc A$\beta$mann and Manfred Bayer.
\newblock Compressive adaptive computational ghost imaging.
\newblock {\em Sci. Rep.}, 3(1):1545, December 2013.

\bibitem{yu2014adaptive}
Wenkai Yu, Mingfei Li, Xuri Yao, Xuefeng Liu, Lingan Wu, and Guangjie Zhai.
\newblock Adaptive compressive ghost imaging based on wavelet trees and sparse
  representation.
\newblock {\em Optics Express}, 22(6):7133--7144, 2014.

\bibitem{rousset2017adaptive}
Florian Rousset, Nicolas Ducros, Andrea Farina, Gianluca Valentini, Cosimo
  Dandrea, and Francoise Peyrin.
\newblock Adaptive basis scan by wavelet prediction for single-pixel imaging.
\newblock {\em IEEE Trans. Comput. Imaging}, 3(1):36--46, 2017.

\bibitem{alemohammad2017high-speed}
Milad Alemohammad, Jasper~R Stroud, Bryan~T Bosworth, and Mark~A Foster.
\newblock High-speed all-optical haar wavelet transform for real-time image
  compression.
\newblock {\em Optics Express}, 25(9):9802--9811, 2017.

\bibitem{strang1993wavelet}
Gilbert Strang.
\newblock Wavelet transforms versus fourier transforms.
\newblock {\em Bulletin of the American Mathematical Society}, 28(2):288--305,
  1993.

\bibitem{strang1996wavelets}
Gilbert Strang and Truong Nguyen.
\newblock {\em Wavelets and filter banks}.
\newblock SIAM, 1996.

\end{thebibliography}

\end{document}